\documentclass[times]{elsarticle}

\usepackage{bsymb}
\usepackage[top=2.5cm, bottom=2.0cm, left=2.5cm, right=2.5cm]{geometry}

\usepackage{cite}
\usepackage{amsmath,amssymb,amsfonts}
\usepackage{algorithmic}
\usepackage{graphicx}

\usepackage{textcomp}
\def\BibTeX{{\rm B\kern-.05em{\sc i\kern-.025em b}\kern-.08em
    T\kern-.1667em\lower.7ex\hbox{E}\kern-.125emX}}

\usepackage{mathtools}

\usepackage[utf8]{inputenc}
\usepackage[T1]{fontenc}
\usepackage{microtype}

\usepackage{color}
\definecolor{gray}{rgb}{0.4,0.4,0.4}
\definecolor{darkblue}{rgb}{0.0,0.0,0.6}
\definecolor{cyan}{rgb}{0.0,0.6,0.6}
\definecolor{keycolor}{rgb}{0,0,0.8}     
\definecolor{labelcolor}{rgb}{0,0.4,0.8} 
\definecolor{codecolor}{rgb}{0,0,0}      
\definecolor{inhcolor}{rgb}{0.6,0.2,0}   
\definecolor{cmtcolor}{rgb}{0,0.4,0}     

\usepackage{bsymb}

\usepackage{listings}

\usepackage{color}
\definecolor{gray}{rgb}{0.4,0.4,0.4}
\definecolor{darkblue}{rgb}{0.0,0.0,1.0}
\definecolor{cyan}{rgb}{0.0,0.6,0.6}

\lstdefinelanguage{MPS}
{
  morestring=[b]",
  morestring=[s]
  ,
  morecomment=[s]{/*}{*/},
  stringstyle=\color{black},
  identifierstyle=\color{black},
  keywordstyle=\color{darkblue},
  morekeywords={domain,model,parent,concepts,concept,is,variable,
  individuals,deduced,attribute,domain,dom,relations,relation,attributes,
  enumerated,elements,
  range,functional,total,maplets,custom,data,sets,set,values,value,type,lexical,form,
  predicates,p1,p2,not}
  ,
  otherkeywords = {=,&,(,),\{,\},>,<,",:,?},
}

\usepackage{lineno,hyperref}
\modulolinenumbers[5]

\journal{Marc Frappier, Régine Laleau, Amel Mammar, Hector Ruiz Barradas}

\usepackage{breqn}
\usepackage{longtable}

\usepackage[linewidth=1pt]{mdframed}
\usepackage{multicol}

\usepackage{flushend}

\begin{document}

\begin{frontmatter}

\title{The Generic SysML/KAOS Domain Metamodel
}


\author{Steve Jeffrey Tueno Fotso and Marc Frappier}
\address{GRIL, Université de Sherbrooke, Sherbrooke, QC J1K 2R1, Canada}
\ead{Steve.Jeffrey.Tueno.Fotso@USherbrooke.ca, Marc.Frappier@USherbrooke.ca}

\author{Régine Laleau}
\address{LACL, Université Paris-Est Créteil, 94010, CRÉTEIL, France}
\ead{laleau@u-pec.fr}

\author{Amel Mammar}
\address{SAMOVAR-CNRS, Télécom SudParis, 91000, Evry, France}
\ead{amel.mammar@telecom-sudparis.eu}

\author{Hector Ruiz Barradas}
\address{ClearSy System Engineering, 13100, Aix-en-Provence, France}
\ead{hector.ruiz-barradas@clearsy.com}


\begin{abstract}
This paper is related to the generalised/generic version of the SysML/KAOS domain metamodel and on translation rules between the new domain models and \textit{B System} specifications.

\end{abstract}


\end{frontmatter}


\section{Background}

\subsection{Event-B and B System} \label{event_b_description_section}
\textit{Event-B } \citep{DBLP:books/daglib/0024570} is an industrial-strength formal method 
for \emph{system modeling}. It is used to
incrementally construct a system specification, using refinement, and to 
 prove useful  properties. 
\textit{B System} is  an  \textit{Event-B} syntactic variant   proposed by \textit{ClearSy}, an industrial partner in the \textit{FORMOSE} project \citep{anr_formose_reference_link}, and supported by   \textit{Atelier B} \citep{clearsy_b_system_link}.
\textit{Event-B} and \textit{B System} have the same semantics defined by proof obligations \citep{DBLP:books/daglib/0024570}.

\begin{figure*}[h]
\begin{center}
\includegraphics[width=1.0\textwidth]{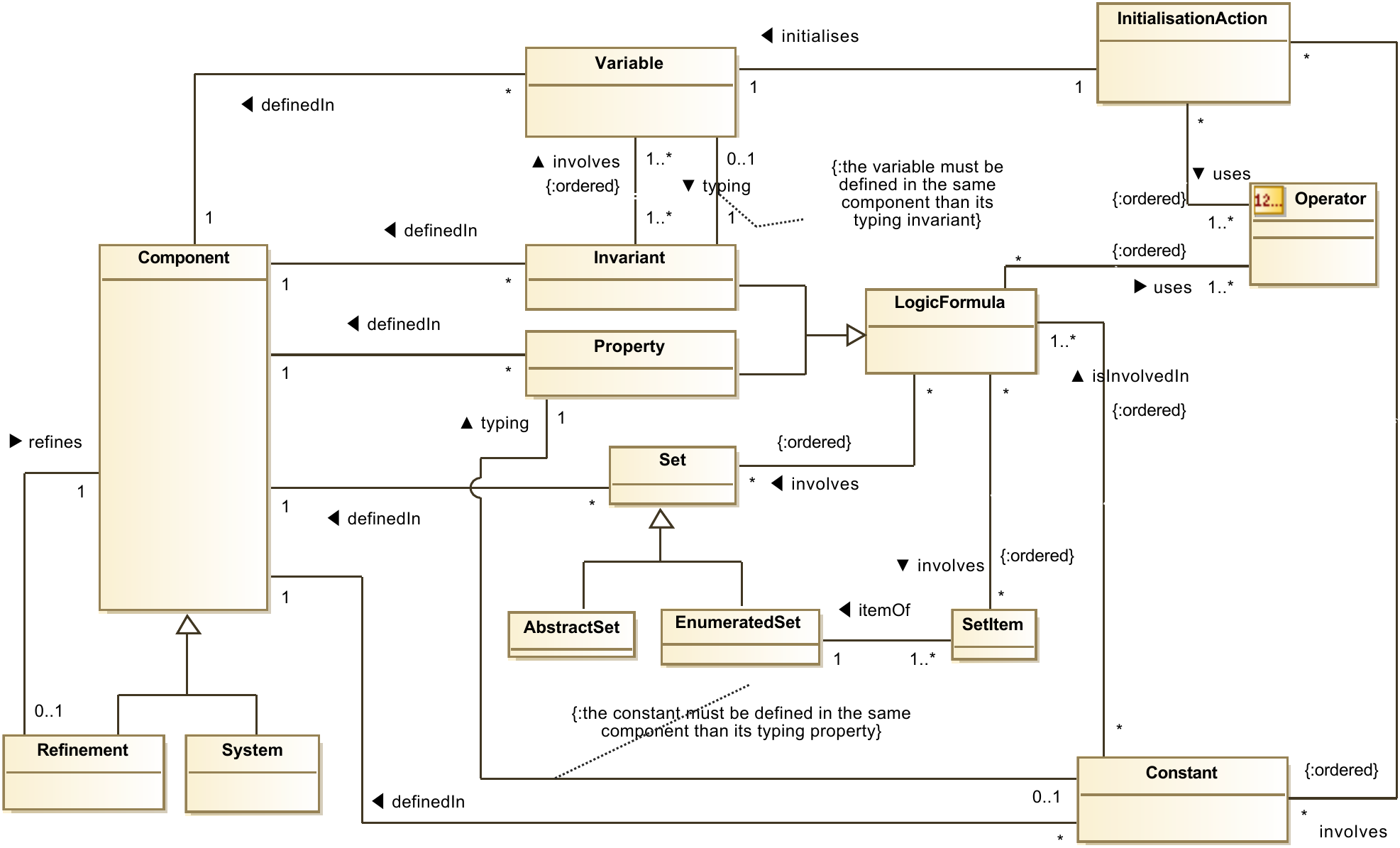}
\end{center}
\caption{\label{eventb_metamodel} Metamodel of the \textit{B System} specification language}
\end{figure*}

 Figure \ref{eventb_metamodel} is 
a metamodel of the \textit{B System} language restricted to  concepts that are relevant to us.
A \textit{B System} specification
consists of components (instances of \textsf{Component}). Each component  can be either a system or a refinement and it   may define static or dynamic elements. A refinement is a  component which refines another one in order  to access the elements defined in it and to reuse them for new constructions. 
Constants, abstract and enumerated sets, and their properties, constitute the static part.
 The dynamic part includes the representation of the  system state   using variables constrained through invariants and initialised through initialisation actions.  Properties and invariants can be categorised as instances of \textsf{LogicFormula}. 
 Variables can be involved only in invariants.
 In our case, it is sufficient to consider  that logic formulas are    successions of operands in relation through operators. Thus,  
  an instance of  \textsf{LogicFormula} references  
its operators (instances of \textsf{Operator}) and its operands that may be instances of \textsf{Variable}, \textsf{Constant},  \textsf{Set} or \textsf{SetItem}. 

\subsection{SysML/KAOS Goal Modeling}

\subsubsection{Presentation}
\textit{SysML/KAOS}~\citep{laleau2010first,DBLP:conf/isola/MammarL16}  is a requirements engineering method
which combines the  traceability  provided by \textit{SysML}~\citep{hause2006sysml}  with  goal expressiveness  provided by KAOS~\citep{DBLP:books/daglib/0025377}.
It  allows the representation of  requirements to be satisfied by a system and of expectations with regards to the environment through a hierarchy of goals.
 The goal hierarchy  is built through a succession of refinements using two main operators: \textbf{\textit{AND}} and  \textbf{\textit{OR}}.
An \textit{\textbf{AND refinement}} decomposes a goal into subgoals, and all of them must be achieved to realise the parent goal.  An \textit{\textbf{OR refinement}} decomposes a goal into subgoals such that the achievement of only one of them is sufficient for the achievement of the parent goal.

For this work, the case study focuses on a communication protocol called \textit{SATURN} proposed by \textit{ClearSy}. SATURN relies on  exchanges of communication frames between different agents connected through a  bus.
This case study is restricted to input/output agents.
Input agents provide boolean data. Each input data undergoes a boolean transformation and the result is made available to  output agents.

\begin{figure}[!h]
\begin{center}
\includegraphics[width=0.8\textwidth]{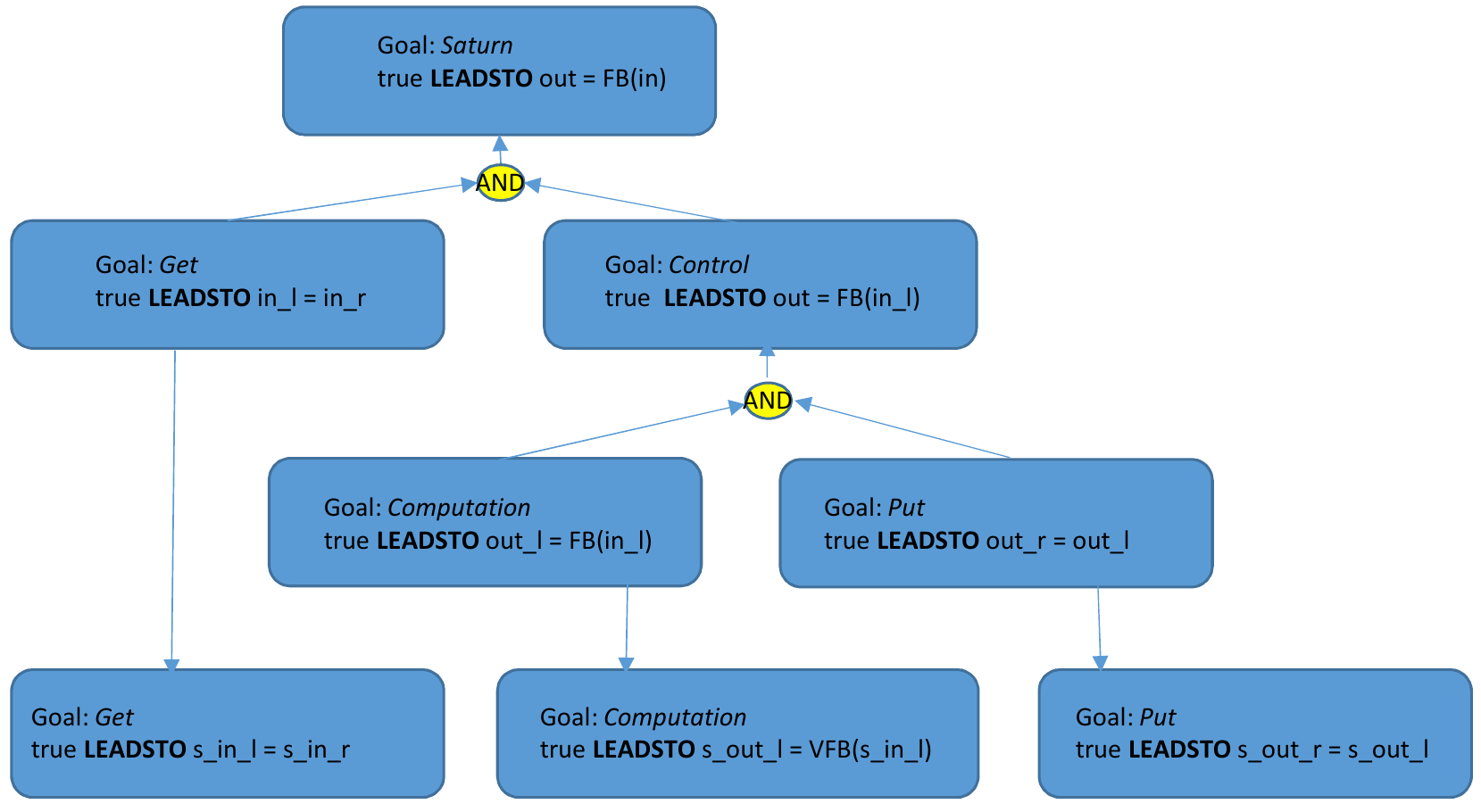}
\end{center}
\caption{\label{saturn_goal_diagram} Excerpt from SATURN system goal diagram}

\end{figure}

Figure \ref{saturn_goal_diagram} is an excerpt from the SysML/KAOS goal diagram representing the functional goals of SATURN. The main purpose
of the system is to transform data provided by input agents (\texttt{in}) and make the result (\texttt{out=FB(in)}) available to output agents.  The purpose gives the root goal \textit{Saturn} of the goal diagram of Fig \ref{saturn_goal_diagram}.
However, goal \textit{Saturn} disregards input reads and result writes.
The \textit{AND} operator is used just after to introduce, at the first refinement level, a goal \textit{Get} for input data acquisition from input agents. Term \texttt{in\_r} designates the data available within input agents and term \texttt{in\_l} designates the input data used to compute the output data.
Similarly,  the second refinement level  introduces a goal \textit{Put}  to make  the result \texttt{out\_l} available to  output agents (\texttt{out\_r} represents the data received by output agents).
The third refinement level refines goals defined within the second refinement level to take into account multiplicities of input and output agents.
Thus, input data acquisition generates a boolean array \texttt{s\_in\_l} instead of 
\texttt{in\_l}, computation becomes a transformation between arrays 
\texttt{s\_out\_l = VFB(s\_in\_l)} and 
result delivery transfers the content of array \texttt{s\_out\_l} to output agents.

 In addition, \textit{SysML/KAOS} includes a domain modeling language which combines the expressiveness of \textit{OWL}~\citep{DBLP:reference/snam/SenguptaH14} and the constraints of  \textit{PLIB}~\citep{DBLP:conf/ifip/Pierra04}.

\subsection{SysML/KAOS Domain Modeling}\label{background_sysmlkaos_domain_language}

\subsubsection{Presentation}
Domain models in SysML/KAOS are represented using ontologies.  These ontologies are expressed using
the SysML/KAOS domain modeling language \citep{1710.00903,sysml_kaos_domain_modeling},   built  
 based on \textit{OWL} \citep{DBLP:reference/snam/SenguptaH14} and \textit{PLIB} \citep{DBLP:conf/ifip/Pierra04}, two well-known and complementary ontology modeling formalisms.

\begin{figure*}[!h]
\begin{center}
\includegraphics[width=1.\textwidth]{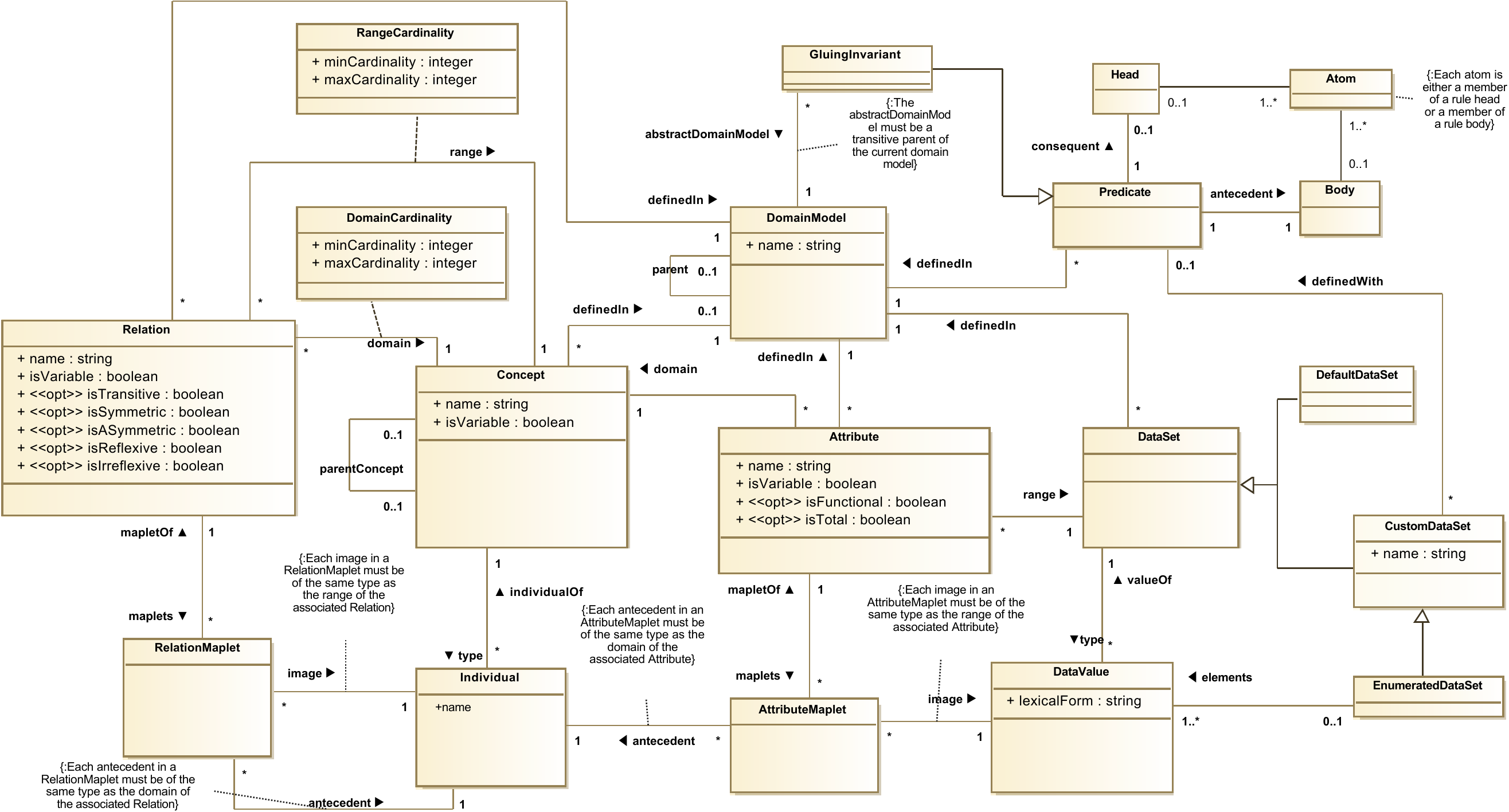}
\end{center}
\caption{\label{our_businessdomain_metamodel} Excerpt from the  metamodel associated with the SysML/KAOS domain modeling language}
\end{figure*}


 Figure \ref{our_businessdomain_metamodel} is  an excerpt from the metamodel associated with the SysML/KAOS  domain modeling language. 
 Each domain model is associated with a level of refinement of the SysML/KAOS goal diagram and is likely to have as its parent, through the \textit{parent} association, another domain model. 
  \textit{Concepts} (instances of  \textsf{Concept})  designate   collections of \textit{individuals} (instances of \textsf{Individual}) with common properties. 
A concept can be declared \textit{variable} (\textit{isVariable=TRUE}) when the set of its individuals can be updated by adding or deleting individuals. Otherwise, it is considered to be \textit{constant} (\textit{isVariable=FALSE}).

\textit{Relations} (instances of \textsf{Relation}) are used to capture  links between concepts, and \textit{attributes} (instances of \textsf{Attribute}) capture  links between concepts and \textit{data sets} (instances of \textsf{DataSet}).
\textit{Relation maplets} (instances of \textsf{RelationMaplet}) capture associations between individuals through  relations and \textit{attribute maplets}  (instances of \textsf{AttributeMaplet}) play the same role for attributes.
A relation   or an attribute can be declared \textit{variable} if the list of maplets related to it  is likely to change over time. Otherwise, it is considered to be \textit{constant}. 
The variability of an association (relation, attribute) is related to  the ability to add or remove  maplets.
 Each \textit{domain cardinality} (instance of \textsf{DomainCardinality}) 
 makes it possible to define, for a relation \texttt{re}, the minimum and maximum limits of the number of individuals of the domain 
  of \texttt{re} 
  that can be put in relation with one
individual of the range of \texttt{re}.
In addition,  the \textit{range cardinality} (instance of \textsf{RangeCardinality})) of \texttt{re}  is used to define similar bounds for the number of individuals of the range of \texttt{re}.

\textit{Predicates} (instances of  \textsf{Predicate})   are used to represent constraints between different elements of the domain model in the form of \textit{horn clauses}: each predicate has a body which represents its \textit{antecedent} and a head which represents its \textit{consequent}, body and head designating conjunctions of atoms.
A data set can be declared abstractly, as a \textit{custom data set} (instance of \textsf{CustomDataSet}), and defined with a predicate.
\textit{Gluing invariants} (instances of \textsf{GluingInvariant}), specialisations of predicates, are used to represent links between data defined within a domain model and those appearing in more abstract domain models,  transitively linked to it through the \textit{parent} association. 
They capture relationships between abstract and concrete data during refinement and are used to discharge  proof obligations. 

\subsubsection{Illustration and Shortcomings}

\begin{figure}[!h]
\begin{center}
\includegraphics[width=0.6\textwidth]{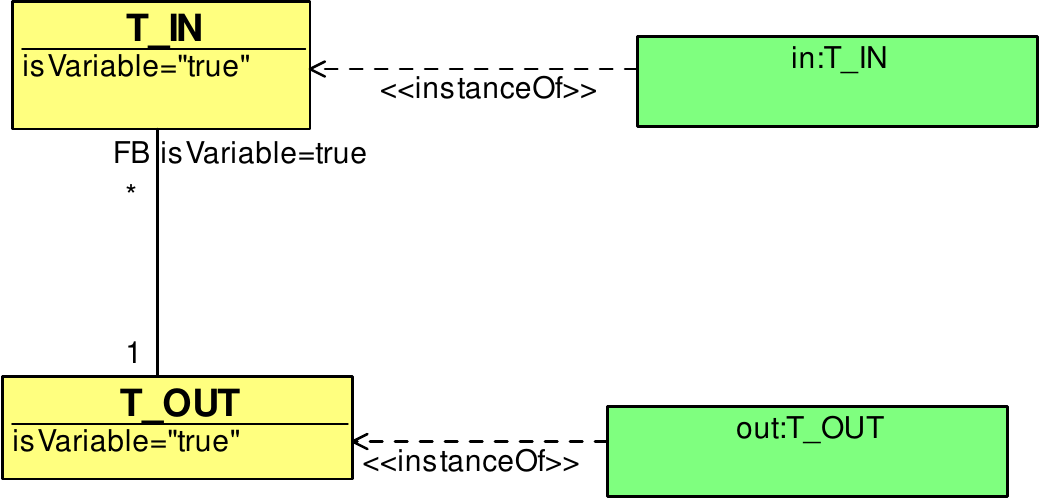}
\caption{\label{saturn_0_cropped} \textit{\textbf{Saturn\_1}}: ontology associated with the root level of the goal diagram of Fig. \ref{saturn_goal_diagram}}
\end{center}
\end{figure}

 Figure \ref{saturn_0_cropped} is an attempt to    represent 
 the domain model  
associated with the root level of the   goal diagram of Fig. \ref{saturn_goal_diagram} using the SysML/KAOS domain modeling language previously described.
It is illustrated using the syntax proposed by   \textit{OWLGred} \citep{owlgred_reference_link}
 and,  for readability purposes, we have decided to hide the representation of  optional characteristics. 
It should be noted that the \textit{individualOf} association is illustrated, through  \textit{OWLGred}, as a stereotyped link with the tag \textit{<<instanceOf>>}.

The type of input data is modeled as a concept \texttt{T\_IN} defining an individual \texttt{in} which represents the input data.
Similarly, the type of output data is modeled as a concept \texttt{T\_OUT} defining an individual \texttt{out} which represents the output data.
The computation function \texttt{FB}  is modeled as a functional relation from \texttt{T\_IN} to \texttt{T\_OUT}.

The first difficulty we encountered is related to the changeability of domain entities.
In fact, the states of  input and output data change dynamically.
In domain model of Fig. \ref{saturn_0_cropped}, a workaround consisted in considering that  concepts \texttt{T\_IN} and \texttt{T\_OUT} and relation 
\texttt{FB} are variables. Thus,  going from a system state where \texttt{out1 = FB(in1)}  to a system state where \texttt{out2 = FB(in2)} is feasible and goes through: (1) withdrawal of maplet \texttt{in1 $\mapsto$ out1} from \texttt{FB}; (2) withdrawal of individual \texttt{in1} from \texttt{T\_IN}; (3) withdrawal of individual \texttt{out1} from \texttt{T\_OUT}; (4) addition of individual \texttt{in2} in \texttt{T\_IN}; (5) addition of individual  \texttt{out2} in \texttt{T\_OUT}; and (6) addition of maplet \texttt{in2 $\mapsto$ out2} in \texttt{FB}.
However, this representation does not conform to SATURN's design.
Indeed, from a conceptual point of view: (1) the input data type must be constant (corresponds to the set of n-tuples of Booleans\footnote{When considering n input agents}); (2) the output data type must be constant (corresponds to the set of m-tuples of Booleans\footnote{When considering m output agents}); (3) the computation function \texttt{FB} is hard-coded and is therefore constant.
What should change are  individuals representing the input and output data.
It is thus necessary to be able to model variable individuals: individual which can dynamically take any value in a given concept. A similar need appears for relations with relation maplets, attributes with attribute maplets and data sets with data values.

Another difficulty has been encountered related to multiplicities of input and output agents (domain model  
associated with the third refinement level of the   goal diagram of Fig. \ref{saturn_goal_diagram}).
Indeed, the array that represents input data needs to be modeled by a relation, ditto for the array that represents output data. Thus, the computation function needs to be modeled by a relation for  which the domain and the range are relations, which is impossible with the current definition of the SysML/KAOS domain modeling language.

 The \textit{SATURN} case study also revealed the need to be able to:
 
 \begin{itemize}
 \item[•] define domain and range cardinalities for attributes;
 \item[•] define a named maplet (instance of \textsf{RelationMaplet} or \textsf{AttributeMaplet})  with or without antecedent and image;
 \item[•] define an initial value for a variable individual, maplet or data value; 
 \item[•] define associations between data sets and maplets between data values;
 \item[•] refine a concept with an association or a data set\footnote{An entity \texttt{ec}, defined in a concrete domain model, refines the  entity \texttt{ea}, defined in an abstract domain model, if it can be deduced that   $ec = ea$ from domain model definitions};
 \item[•] refine an individual with a maplet or a data value.
 \end{itemize}
 
 We have therefore identified the need to build a generalisation of the metamodel of Fig. \ref{our_businessdomain_metamodel} which enriches the expressiveness of the SysML/KAOS domain modeling language while preserving the fundamental constraints identified in \citep{1710.00903,sysml_kaos_domain_modeling}.
 A major contribution of this new metamodel is that it federates notions of concept, data set, attribute and relation as well as notions of individual, maplet and data value that have always been considered distinct by ontology modeling languages.
 Additional constraints are defined to preserve the formal semantics of the language and to ensure unambiguous transformation of any domain model to a \textit{B System} specification.

\section{The New SysML/KAOS Domain Modeling Language}

\subsection{Presentation}

\begin{figure*}[!h]
\begin{center}
\includegraphics[width=1.0\textwidth]{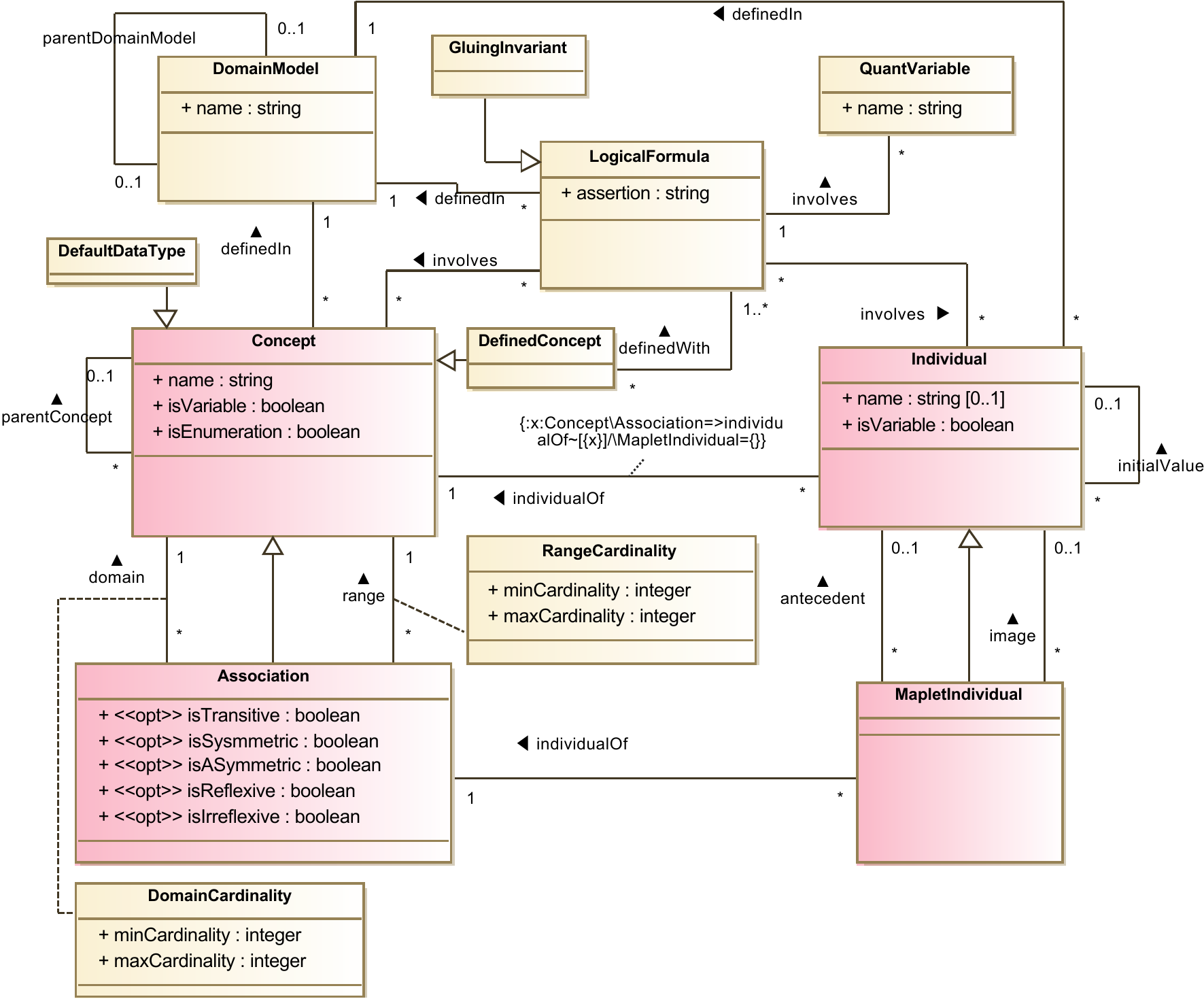}
\end{center}
\caption{\label{our_new_businessdomain_metamodel} Excerpt from the updated SysML/KAOS domain  metamodel}
\end{figure*}

 Figure \ref{our_new_businessdomain_metamodel} is  an excerpt from the updated SysML/KAOS domain  metamodel. 

\subsubsection{Description} 
 
Domain models are also associated with  levels of refinement of the SysML/KAOS goal model. 
  \textit{Concepts} (instances of  \textsf{Concept})  designate   collections of \textit{individuals} (instances of \textsf{Individual}) with common properties. 
A concept can be declared \textit{variable} (\textit{isVariable=TRUE}) when the set of its individuals can be updated by adding or deleting individuals. Otherwise, it is considered to be \textit{constant} (\textit{isVariable=FALSE}).
In addition, a concept can be an enumeration (\textit{isEnumeration=TRUE}) if all its individuals are defined within the domain model.
It should be noted that an individual can be \textit{variable} (\textit{isVariable=TRUE}) if it is introduced to represent a system state variable: it can represent different individuals at different system states. Otherwise, it is  \textit{constant} (\textit{isVariable=FALSE}).

\textit{Associations} (instances of \textsf{Association}) are concepts used to capture  links between concepts.
\textit{Maplet individuals} (instances of \textsf{MapletIndividual}) capture associations between individuals through  associations. 
Each named maplet individual can reference an antecedent and an image. When the maplet individual is unnamed, the antecedent and the image must be specified. 
The variability of an association is related to  the ability to add or remove  maplets.
 Each \textit{domain cardinality} (instance of \textsf{DomainCardinality}) 
 makes it possible to define, for an association \texttt{re}, the minimum and maximum limits of the number of individuals of the domain 
  of \texttt{re} 
  that can be put in relation with one
individual of the range of \texttt{re}.
In addition,  the \textit{range cardinality} (instance of \textsf{RangeCardinality})) of \texttt{re}  is used to define similar bounds for the number of individuals of the range of \texttt{re}.

Class \textsf{LogicalFormula} replaces class \textsf{Predicate} of the metamodel of Fig. \ref{our_businessdomain_metamodel} to represent constraints between domain model  elements.

 
 \subsubsection{Additional Constraints}
This section defines the  constraints that are required to preserve the formal semantics of the domain modeling language and to ensure an unambiguous transformation of any domain model to a \textit{B System} specification. The constraints are defined using the \textit{B} syntax \citep{DBLP:books/daglib/0024570}.

\begin{itemize}
\item[•] $x \in \textsf{Concept} \setminus \textsf{Association} \\\Rightarrow \textsf{Individual\_individualOf\_Concept}^{-1}[\{x\}]\cap \textsf{MapletIndividual}= \emptyset$: if concept \texttt{x} is not an association, then no individual of \texttt{x} can be a maplet individual.
\item[•] $x \in \textsf{MapletIndividual} \cap dom(\textsf{MapletIndividual\_antecedent\_Individual}) \\\Rightarrow \textsf{MapletIndividual\_antecedent\_Individual}(x) \in \textsf{Association\_domain\_Concept}(\textsf{Individual\_individualOf\_Concept}(x))$: if maplet individual  \texttt{x} has an antecedent, then the antecedent is an individual of the domain of its association.
\item[•] $x \in \textsf{MapletIndividual} \cap dom(\textsf{MapletIndividual\_image\_Individual}) \\\Rightarrow \textsf{MapletIndividual\_image\_Individual}(x) \in \textsf{Association\_range\_Concept}(\textsf{Individual\_individualOf\_Concept}(x))$: if maplet individual  \texttt{x} has an image, then the image is an individual of the range of its association.
\item[•] $ind \in \textsf{Individual}\setminus \textsf{MapletIndividual} \Rightarrow ind \in dom(\textsf{Individual\_name})$: every individual which is not a maplet individual must be named.
\item[•] $ind \in \textsf{Individual}\setminus dom(\textsf{Individual\_name}) \Rightarrow \textsf{Individual\_isVariable}(ind) = FALSE$: every unnamed individual must be constant.
\item[•] $ind \in \textsf{MapletIndividual} \cap dom(\textsf{MapletIndividual\_antecedent\_Individual})\cap dom(\textsf{MapletIndividual\_image\_Individual}) \\\Rightarrow  (\textsf{MapletIndividual\_antecedent\_Individual}(ind) \in dom(\textsf{Individual\_name}) \wedge \textsf{MapletIndividual\_image\_Individual}(ind) \in dom(\textsf{Individual\_name}))$: antecedents and images of maplet individuals must be named.
\item[•] $ind \in \textsf{MapletIndividual}\setminus dom(\textsf{Individual\_name}) \\\Rightarrow  ind \in dom(\textsf{MapletIndividual\_antecedent\_Individual}) \cap dom(\textsf{MapletIndividual\_image\_Individual})$: every unnamed maplet individual must have an antecedent and an image.
\item[•] $x \in \textsf{Concept} \setminus (\textsf{Association} \cup \textsf{DefinedConcept} \cup   dom(\textsf{Concept\_parent\_Concept}))\\ \Rightarrow \textsf{Concept\_isVariable}(x)=FALSE$: every abstract concept (that has no parent concept) that is not an association  must be constant.
\item[•] $x \in  \textsf{Concept} \wedge \textsf{Concept\_isEnumeration}(x)=TRUE \Rightarrow \textsf{Concept\_isVariable}(x)=FALSE$: every concept that is an enumeration must be constant.
\item[•] $(ind \in \textsf{MapletIndividual} \cap dom(\textsf{MapletIndividual\_antecedent\_Individual})\cap dom(\textsf{MapletIndividual\_image\_Individual}) \\\wedge \textsf{Individual\_isVariable}(ind) = FALSE) \\\Rightarrow (\textsf{Individual\_isVariable}(\textsf{MapletIndividual\_antecedent\_Individual}(ind))= FALSE \\\wedge \textsf{Individual\_isVariable}(\textsf{MapletIndividual\_image\_Individual}(ind))= FALSE)$: antecedents and images of constant maplet individuals must be constant.
\item[•] $(x \in \textsf{Association} \wedge \textsf{Concept\_isVariable}(x) = FALSE) \\\Rightarrow (\textsf{Concept\_isVariable}(\textsf{Association\_domain\_Concept}(x))= FALSE \\\wedge \textsf{Concept\_isVariable}(\textsf{Association\_range\_Concept}(x))= FALSE)$: domains and ranges of constant associations must be constant.
\end{itemize}

\subsection{Illustration}
\begin{figure}[!h]
\begin{center}
\includegraphics[width=0.5\textwidth]{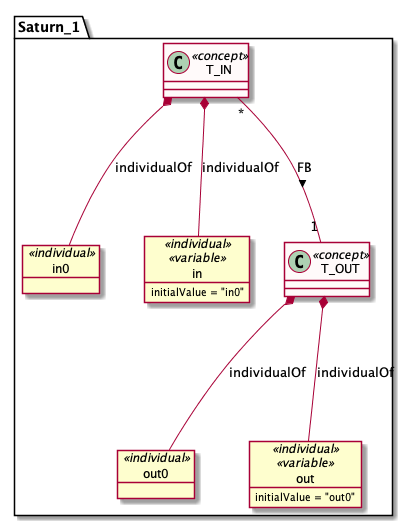}
\caption{\label{Saturn_1} \textit{\textbf{Saturn\_1}}: ontology associated with the root level of the goal diagram of Fig. \ref{saturn_goal_diagram}}
\end{center}
\end{figure}

\begin{figure}[!h]
\begin{center}
\includegraphics[width=0.5\textwidth]{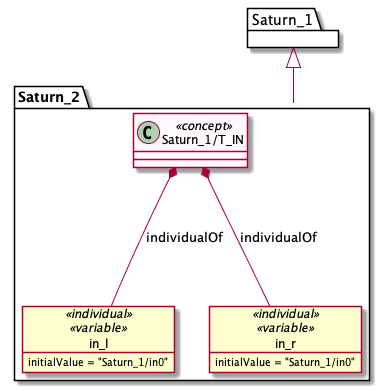}
\caption{\label{Saturn_2} \textit{\textbf{Saturn\_2}}: ontology associated with the first refinement level of the goal diagram of Fig. \ref{saturn_goal_diagram}}
\end{center}
\end{figure}

\begin{figure}[!h]
\begin{center}
\includegraphics[width=0.5\textwidth]{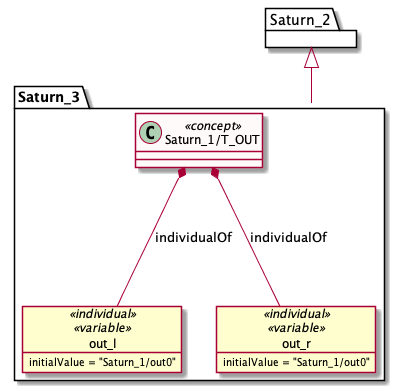}
\caption{\label{Saturn_3} \textit{\textbf{Saturn\_3}}: ontology associated with the second refinement level of the goal diagram of Fig. \ref{saturn_goal_diagram}}
\end{center}
\end{figure}

\begin{figure}[!h]
\begin{center}
\includegraphics[width=1.\textwidth]{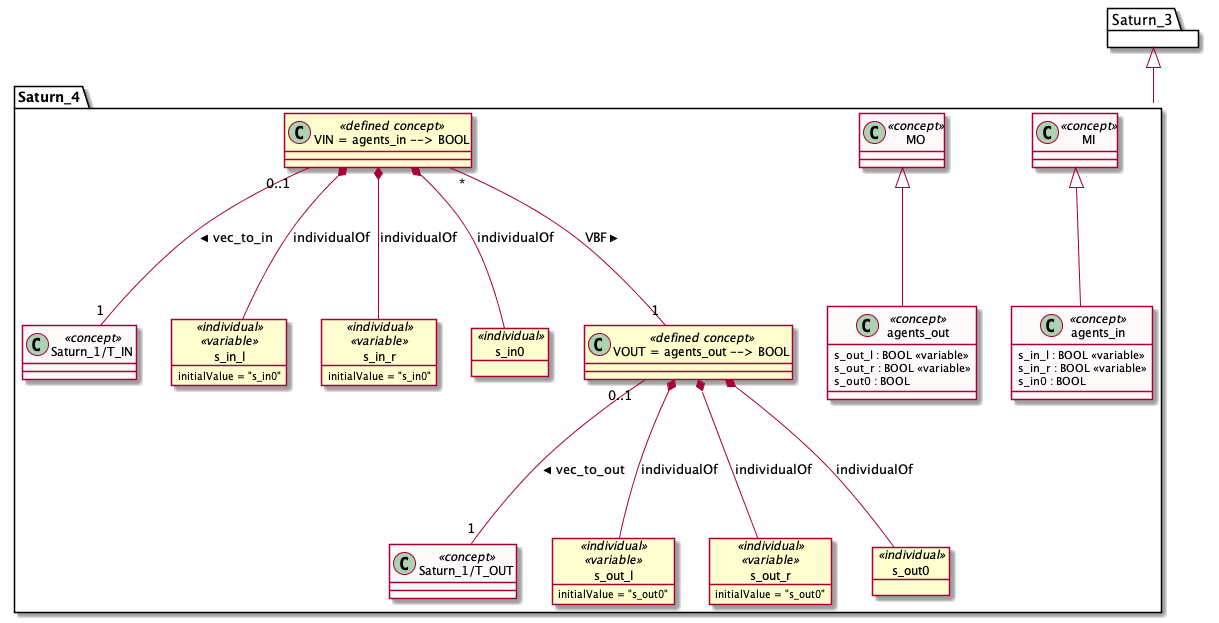}
\caption{\label{Saturn_4} \textit{\textbf{Saturn\_4}}: ontology associated with the third refinement level of the goal diagram of Fig. \ref{saturn_goal_diagram}}
\end{center}
\end{figure}

 Figures \ref{Saturn_1}, \ref{Saturn_2}, \ref{Saturn_3}, \ref{Saturn_4} 
    represent 
  domain models  
associated with refinement
levels 0 (root level) $\upto$ 3  of the   goal diagram of Fig. \ref{saturn_goal_diagram} using the updated SysML/KAOS domain modeling language.
They are illustrated using the syntax proposed by   the \textit{SysML/KAOS Domain Modeling tool}  \citep{SysML_KAOS_Domain_Model_Parser_link}\footnote{The tool has been implemented on top of \textit{Jetbrains MPS} \citep{jetbrains_mps} and PlantUML \citep{roques2015plantuml} to provide a proof of concept of the SysML/KAOS Domain Modeling Language.}
 and,  for readability purposes, we have decided to hide the representation of  optional characteristics.

 In domain model \texttt{Saturn\_1} (Fig. \ref{Saturn_1}), the type of input data is modeled as a constant concept \texttt{T\_IN} (instance of class \textsf{Concept} of Fig. \ref{our_new_businessdomain_metamodel}) defining a variable individual \texttt{in} (instance of class \textsf{Individual} of Fig. \ref{our_new_businessdomain_metamodel}) which represents the input data.
Similarly, the type of output data is modeled as a constant concept \texttt{T\_OUT} defining a variable individual \texttt{out} which represents the output data.
Finally, the computation function \texttt{FB}  is modeled as a functional association (instance of class \textsf{Association} of Fig. \ref{our_new_businessdomain_metamodel}) from \texttt{T\_IN} to \texttt{T\_OUT}.
Constant individuals \texttt{in0} and \texttt{out0} represent respectively the initial value of  \texttt{in} and that of \texttt{out}.

 In domain model \texttt{Saturn\_2} (Fig. \ref{Saturn_2}) which refines \texttt{Saturn\_1}, individual \texttt{in} is refined by an individual named \texttt{in\_l} ($in\_l = in$) and a new variable individual named \texttt{in\_r} is defined to represent the acquired input data. Similarly, in domain model \texttt{Saturn\_3} (domain model associated with refinement
level 2  of the   goal diagram of Fig. \ref{saturn_goal_diagram}), \texttt{out} is refined by \texttt{out\_l} ($out\_l = out$) and individual \texttt{out\_r} is added.
 
 In domain model \texttt{Saturn\_4} (Fig. \ref{Saturn_4}) which refines \texttt{Saturn\_3}, two concepts are defined: \texttt{MI} which represents the set of input agents and \texttt{MO} which represents the set of output agents. Concept \texttt{agents\_in} (respectively \texttt{agents\_out})  is a subconcept of \texttt{MI} (respectively \texttt{MO})  which represents the set of input (respectively output) agents that are active. 
 Concept \texttt{VIN}, defined as the set of total functions from \texttt{agents\_in} to \texttt{BOOL} ($VIN=agents\_in \longrightarrow BOOL$ where $BOOL=\{TRUE, FALSE\}$), represents the type of  input data which are now arrays. Similarly, concept \texttt{VOUT} ($VOUT=agents\_out \longrightarrow BOOL$)  represents the type of output data. 
 Individuals \texttt{in\_l}, \texttt{in\_r}, \texttt{out\_l} and \texttt{out\_r} are refined respectively by  individuals   \texttt{s\_in\_l}, \texttt{s\_in\_r}, \texttt{s\_out\_l} and \texttt{s\_out\_r} using total injective associations \texttt{vec\_to\_in} from \texttt{VIN} to \texttt{T\_IN} and \texttt{vec\_to\_out} from \texttt{VOUT} to \texttt{T\_OUT}: $in\_l = vec\_to\_in(s\_in\_l), in\_r = vec\_to\_in(s\_in\_r), out\_l = vec\_to\_out(s\_out\_l), out\_r = vec\_to\_out(s\_out\_r)$.
 Finally, the computation function   is modeled as a functional association named \texttt{VBF} from \texttt{VIN} to \texttt{VOUT}: $VBF = vec\_to\_in; FB; vec\_to\_out^{-1}$ (operator $;$ is the association composition operator used in logical formula assertions).

\section{Updates in Translation Rules from Domain Models to B System Specifications}
In the following, we describe a set of rules that allow to obtain a \textit{B System} specification from domain models that conform to the updated SysML/KAOS domain modeling language. 

Table \ref{tableau_recapitulatif_correspondances} gives the  translation rules.
It should be noted that \textit{o\_x} designates the result of the translation of \textit{x}. In addition, 
when used, qualifier \textit{abstract}  denotes "without parent".
The rules have been implemented within the \textit{SysML/KAOS Domain Modeling tool}  \citep{SysML_KAOS_Domain_Model_Parser_link} built on top of \textit{Jetbrains MPS} \citep{jetbrains_mps} and PlantUML \citep{roques2015plantuml} to provide a proof of concept of the   SysML/KAOS Domain Modeling Language.
Rules \textit{3, 4, 6$\upto$8}, and \textit{12$\upto$16} have undergone  significant updates
 to the previously defined translation rules
\citep{submit/2184994}.

\begin{scriptsize}
\begin{center}
\begin{longtable}{|p{.1cm}|p{2.2cm}|p{0.9cm}|p{6.cm}|p{0.9cm}|p{5.7cm}|}
\caption{\label{tableau_recapitulatif_correspondances} The translation rules}\\
\hline
& &\multicolumn{2}{|c|}{\textbf{Domain Model}} & \multicolumn{2}{|c|}{\textbf{B System}} \\
\hline
& \textbf{Translation Of} &\textbf{Element} & \textbf{Constraint} & \textbf{Element} & \textbf{Constraint} \\
\hline
1 & \textbf{Abstract domain model}  & DM & $DM \in \textsf{DomainModel}$ \newline $DM \notin dom(\textsf{DomainModel\_parent\_DomainModel})$ & o\_DM & $o\_DM \in \textsf{System}$ \\
\hline
2 & \textbf{Domain model with parent} & DM PDM & $\{DM,PDM\} \subseteq \textsf{DomainModel} $ \newline
$PDM = \textsf{DomainModel\_parent\_DomainModel}(DM)$ \newline
$o\_PDM \in \textsf{Component}$ 
  & o\_DM & $o\_DM \in \textsf{Refinement} $ \newline
  $\textsf{Refinement\_refines\_Component}(o\_DM) = o\_PDM$
 \\
\hline
3 & \textbf{Abstract concept that is not an enumeration} & CO & $CO \in \textsf{Concept}\setminus (\textsf{Association}\cup \textsf{DefinedConcept} \cup \textsf{DefaultDataType}) $ 
\newline\newline $CO \notin dom(\textsf{Concept\_parent\_Concept})$  
\newline\newline  $\textsf{Concept\_isEnumeration}(CO) =  FALSE$
& o\_CO & 
$o\_CO \in  \textsf{AbstractSet}$
\\
\hline
4 & \textbf{Abstract concept that is  an enumeration} & CO $(I_j)_{j \in 1..n}$ & $CO \in \textsf{Concept}\setminus (\textsf{Association}\cup \textsf{DefinedConcept} \cup \textsf{DefaultDataType}) $ 
\newline\newline $CO \notin dom(\textsf{Concept\_parent\_Concept})$  
\newline\newline  $\textsf{Concept\_isEnumeration}(CO) =  TRUE$
\newline\newline $\forall j \in 1..n, I_j \in \textsf{Individual} \newline \wedge \textsf{Individual\_individualOf\_Concept}(I_j) = CO \newline\wedge \textsf{Individual\_isVariable}(I_j)=FALSE$
& o\_CO $(o\_I_j)_{j \in 1..n}$ & 
$o\_CO \in  \textsf{EnumeratedSet}$
\newline\newline $\forall j \in 1..n, o\_I_j \in \textsf{SetItem} \newline \wedge \textsf{SetItem\_itemOf\_EnumeratedSet}(o\_I_j) = o\_CO$
\\
\hline
5 & \textbf{Concept with constant parent} & CO PCO 
&
$\{CO,PCO\} \subseteq \textsf{Concept}$ 
\newline$\textsf{Concept\_parent\_Concept}(CO) = PCO$  
\newline$o\_PCO \in \textsf{Set} \cup \textsf{Constant}$
& o\_CO &
\textbf{IF} $Concept\_isVariable(CO) =  FALSE$
 \newline \textbf{THEN} $o\_CO \in  \textsf{Constant}$
  \newline \textbf{ELSE} $o\_CO \in  \textsf{Variable}$
\newline \textbf{\textsf{LogicFormula}:} $o\_CO \subseteq o\_PCO$
\\
\hline
6 & \textbf{Constant concept with variable parent} & CO PCO PPCO 
&
$\{CO,PCO, PPCO\} \subseteq \textsf{Concept}$
\newline $\textsf{Concept\_isVariable}(CO) =  FALSE$  
\newline  $\textsf{Concept\_parent\_Concept}(CO) = PCO$  
\newline  $o\_PCO \in \textsf{Variable}$
\newline $PPCO \in  (closure1(\textsf{Concept\_parent\_Concept}))[\{PCO\}]$\footnote{\textit{closure1(\textsf{Concept\_parent\_Concept})} designates the transitive closure of  relation \textsf{\textsf{Concept\_parent\_Concept}}}
\newline  $o\_PPCO \in \textsf{Set} \cup \textsf{Constant}$
& o\_CO &
$o\_CO \in  \textsf{Constant}$
\newline \textbf{\textsf{Property}:} $o\_CO \subseteq o\_PPCO$
\newline \textbf{\textsf{Invariant}:} $o\_CO \subseteq o\_PCO$
\\
\hline
7 & \textbf{Variable concept with variable parent} & CO PCO 
&
$\{CO,PCO\} \subseteq \textsf{Concept}$
\newline  $\textsf{Concept\_isVariable}(CO) =  TRUE$  
\newline  $\textsf{Concept\_parent\_Concept}(CO) = PCO$  
\newline $o\_PCO \in \textsf{Variable}$
& o\_CO &
$o\_CO \in  \textsf{Variable}$
\newline \textbf{\textsf{Invariant}:} $o\_CO \subseteq o\_PCO$
\\
\hline
8 & \textbf{Enumerated concept with parent} & CO $(I_j)_{j \in 1..n}$
&
$CO \in dom(\textsf{\textsf{Concept\_parent\_Concept}})$
\newline\newline  $\textsf{Concept\_isEnumeration}(CO) =  TRUE$
\newline\newline $\forall j \in 1..n, I_j \in \textsf{Individual} \newline \wedge \textsf{Individual\_individualOf\_Concept}(I_j) = CO \newline \wedge \textsf{Individual\_isVariable}(I_j)=FALSE$
\newline\newline $o\_CO \in  \textsf{Constant}$\footnote{Every concrete enumeration is a constant}
\newline\newline $\forall j \in 1..n, o\_I_j \in  o\_CO$
&  &
 \textbf{\textsf{Property}:} $o\_CO = (o\_I_j)_{j \in 1..n}$
\\
\hline
(9) & \textbf{Association or defined concept  without parent} & CO  
&
$CO \in  (\textsf{DefinedConcept} \cup \textsf{Association})$
\newline  $CO \notin dom(\textsf{Concept\_parent\_Concept})$\footnote{If $CO$ has a parent concept, $o\_CO$ must be introduced by rule 5.  It is therefore necessary to ensure that this is not the case.} 
\newline

  To ensure that each variable or constant is typed, this rule has to be combined with either rule 10, for associations, or with a translation of the defining logical formula (contained in \textsf{definedWith}), for defined concepts.
& o\_CO &
\textbf{IF} $\textsf{Concept\_isVariable}(CO) =  FALSE$
 \newline \textbf{THEN} $o\_CO \in  \textsf{Constant}$
  \newline \textbf{ELSE} $o\_CO \in  \textsf{Variable}$
\\
\hline
10 & \textbf{Association} &
AS 
CO1 
CO2 
da di ra ri
& $\{CO1,CO2\} \subseteq \textsf{Concept} $ 
 \newline  $AS \in \textsf{Association}$
  \newline  $CO1 = \textsf{Association\_domain\_Concept}(AS)$
  \newline  $CO2 = \textsf{Association\_range\_Concept}(AS)$
  \newline  $\textsf{Association\_DomainCardinality\_maxCardinality}(AS)=da$
    \newline  $\textsf{Association\_DomainCardinality\_minCardinality}(AS)=di$
        \newline  $\textsf{Association\_RangeCardinality\_maxCardinality}(AS)=ra$
            \newline  $\textsf{Association\_RangeCardinality\_minCardinality}(AS)=ri$
            \newline  $o\_AS \in \textsf{Constant} \cup \textsf{Variable}$
  \newline  $\{o\_CO1,o\_CO2\} \subseteq (\textsf{Set} \cup \textsf{Constant} \cup \textsf{Variable})$ 
 & 
 T\_o\_AS
 &  
 \textbf{IF} $\textsf{Concept\_isVariable}(CO1) =  FALSE \newline \wedge \textsf{Concept\_isVariable}(CO2) =  FALSE$
 \newline \textbf{THEN} $T\_o\_AS \in  \textsf{Constant}$
  \newline \textbf{ELSE} $T\_o\_AS \in  \textsf{Variable}$
  \newline
  
 \textbf{IF}  $\{ra,ri,da,di\} = \{1\}$ \newline \textbf{THEN} \textbf{\textsf{LogicFormula}:}  $T\_o\_AS =  o\_CO1 \tbij o\_CO2$
    \newline \textbf{ELSE IF}  $\{ra,ri,da\} = \{1\}$ \newline  \textbf{THEN} \textbf{\textsf{LogicFormula}:}  $T\_o\_AS =   o\_CO1 \tinj o\_CO2$
      \newline \textbf{ELSE IF}  $\{ra,ri,di\} = \{1\}$ \newline \textbf{THEN} \textbf{\textsf{LogicFormula}:}  $T\_o\_AS =  o\_CO1 \tsur o\_CO2$
     \begin{scriptsize}
      \newline \textbf{ELSE IF}  $\{ra,di\} = \{1\}$ \newline \textbf{THEN} \textbf{\textsf{LogicFormula}:}  $T\_o\_AS =  o\_CO1 \psur o\_CO2$
       \newline \textbf{ELSE IF}  $\{ra,da\} = \{1\}$ \newline \textbf{THEN} \textbf{\textsf{LogicFormula}:}  $T\_o\_AS =  o\_CO1 \pinj o\_CO2$
   \newline \textbf{ELSE IF}  $\{ra,ri\} = \{1\}$ \newline \textbf{THEN} \textbf{\textsf{LogicFormula}:}  $T\_o\_AS =  o\_CO1 \longrightarrow o\_CO2$
  \newline \textbf{ELSE IF}  $ra=1$  
  \newline \textbf{THEN}   \textbf{\textsf{LogicFormula}:}  $T\_o\_AS =  o\_CO1 \pfun o\_CO2$
     \end{scriptsize}
    \newline \textbf{ELSE} 
	\newline \hspace*{0.20in} \textbf{\textsf{LogicFormula}:}  $T\_o\_AS =  o\_CO1 \leftrightarrow o\_CO2$ 
    	\newline \hspace*{0.20in}	$\wedge \forall x. (x \in CO2 \Rightarrow card(o\_RE^{-1}[\{x\}])  \in di..da)$
    	\newline \hspace*{0.20in} $\wedge \forall x. (x \in CO1 \Rightarrow card(o\_RE[\{x\}])  \in ri..ra)$
    	\newline
    	
    	\textbf{\textsf{LogicFormula}:}  $o\_AS \in T\_o\_AS$
 \\
\hline
11 & \textbf{Individual of a constant concept that is not an abstract enumeration} &
Ind CO & $Ind \in \textsf{Individual} \setminus \textsf{MapletIndividual}$ \newline
 $CO = \textsf{Individual\_individualOf\_Concept}(Ind)$   \newline
  $o\_CO \in \textsf{AbstractSet} \cup  \textsf{Constant}$
& o\_Ind & 
\textbf{IF}  $\textsf{Individual\_isVariable}(Ind) = TRUE$
 \newline \textbf{THEN} $o\_Ind \in  \textsf{Variable}$
  \newline \textbf{ELSE} $o\_Ind \in  \textsf{Constant}$
\newline \textbf{\textsf{LogicFormula}:} $o\_Ind \in o\_CO$
\\
\hline
12 & \textbf{Constant individual of a variable concept} &
Ind CO PPCO & $Ind \in \textsf{Individual} \setminus \textsf{MapletIndividual}$ \newline
$\textsf{Individual\_isVariable}(Ind) = FALSE$ \newline
 $CO = \textsf{Individual\_individualOf\_Concept}(Ind)$   \newline
  $o\_CO \in   \textsf{Variable}$ \newline
   $PPCO \in \textsf{Concept}$ \newline
   $PPCO \in  (closure1(\textsf{Concept\_parent\_Concept}))[\{CO\}]$ 
    \newline $o\_PPCO \in \textsf{Set} \cup \textsf{Constant}$
& o\_Ind & 
$o\_Ind \in  \textsf{Constant}$
\newline \textbf{\textsf{Property}:} $o\_Ind \in o\_PPCO$
\newline \textbf{\textsf{Invariant}:} $o\_Ind \in o\_CO$
\\
\hline
13 & \textbf{Variable individual of a variable concept} &
Ind CO & $Ind \in \textsf{Individual} \setminus \textsf{MapletIndividual}$ \newline
$\textsf{Individual\_isVariable}(Ind) = TRUE$ \newline
 $CO = \textsf{Individual\_individualOf\_Concept}(Ind)$   \newline
  $o\_CO \in   \textsf{Variable}$ \newline
& o\_Ind & 
$o\_Ind \in  \textsf{Variable}$
\newline \textbf{\textsf{Invariant}:} $o\_Ind \in o\_CO$
\\
\hline
14 & \textbf{Variable individual of a concept that is an abstract enumeration} &
Ind CO & $Ind \in \textsf{Individual} \setminus \textsf{MapletIndividual}$ \newline
$\textsf{Individual\_isVariable}(Ind)=TRUE$ \newline
 $CO = \textsf{Individual\_individualOf\_Concept}(Ind)$   \newline
  $\textsf{Concept\_isEnumeration}(CO) = TRUE$   \newline
  $CO \notin dom(\textsf{Concept\_parent\_Concept})$   \newline
  $o\_CO \in \textsf{EnumeratedSet}$
& o\_Ind & 
 $o\_Ind \in  \textsf{Variable}$
\newline \textbf{\textsf{Invariant}:} $o\_Ind \in o\_CO$
\\
\hline
15 & \textbf{Maplet individual} &
Ind AS Ant Im  PPCO1 PPCO2
& $Ind \in  \textsf{MapletIndividual}$ 
 \newline\newline $AS = \textsf{Individual\_individualOf\_Concept}(Ind)$\footnote{\textit{AS} must be an association}   
  \newline\newline $o\_AS \in  \textsf{Constant} \cup \textsf{Variable}$
  \newline\newline $Ind \in dom(\textsf{MapletIndividual\_antecedent\_Individual}) \newline \Rightarrow Ant = \textsf{MapletIndividual\_antecedent\_Individual}(Ind)$
   \newline\newline $o\_Ant \in  \textsf{Constant} \cup \textsf{Variable}$
     \newline\newline $Ind \in dom(\textsf{MapletIndividual\_image\_Individual}) \newline \Rightarrow Im = \textsf{MapletIndividual\_image\_Individual}(Ind)$
   \newline\newline $o\_Im \in  \textsf{Constant} \cup \textsf{Variable}$
   \newline\newline $\{PPCO1, PPCO2\} \subseteq \textsf{Concept}$ 
   \newline\newline $PPCO1 \in  (closure1(\textsf{Concept\_parent\_}$ 
   \newline $\textsf{Concept}))[\{\textsf{Association\_domain\_Concept}(AS)\}]$ 
      \newline\newline $PPCO2 \in  (closure1(\textsf{Concept\_parent\_}$ 
   \newline $\textsf{Concept}))[\{\textsf{MapletIndividual\_range\_Individual}(AS)\}]$ 
    \newline\newline $\{o\_PPCO1, o\_PPCO2\} \subseteq \textsf{Set} \cup \textsf{Constant}$
& o\_Ind & 
\textbf{IF}  $Ind \in dom(\textsf{Individual\_name})$
\newline \textbf{THEN}
\newline \hspace*{0.05in} \textbf{IF}  $\textsf{Individual\_isVariable}(Ind) = TRUE$
 \newline \hspace*{0.05in} \textbf{THEN} 
 	\newline \hspace*{0.1in} $o\_Ind \in  \textsf{Variable}$
 	\newline \hspace*{0.1in} \textbf{\textsf{Invariant}:} $o\_Ind \in o\_AS$
 	\newline \hspace*{0.10in} \textbf{IF} $Ind \in dom(\textsf{MapletIndividual\_antecedent\_Individual}) \newline\hspace*{0.20in} \cap dom(\textsf{MapletIndividual\_image\_Individual})$
 	\newline \hspace*{0.10in} \textbf{THEN} \textbf{\textsf{Invariant}:} $o\_Ind = o\_Ant \mapsto o\_Im$
  \newline \hspace*{0.05in}  \textbf{ELSE} 
  \newline \hspace*{0.10in} $o\_Ind \in  \textsf{Constant}$
  \newline \hspace*{0.10in} \textbf{IF} $o\_AS \in  \textsf{Constant}$
 	\newline \hspace*{0.10in} \textbf{THEN} \textbf{\textsf{Property}:} $o\_Ind \in o\_AS$
 	\newline \hspace*{0.10in} \textbf{ELSE} 
 	 	\newline \hspace*{0.15in} \textbf{\textsf{Property}:} $o\_Ind \in o\_PPCO1 \leftrightarrow o\_PPCO2$
 	 	\newline \hspace*{0.15in} \textbf{\textsf{Invariant}:} $o\_Ind \in o\_AS$
\newline \hspace*{0.10in} \textbf{IF} $Ind \in dom(\textsf{MapletIndividual\_antecedent\_Individual}) \newline\hspace*{0.20in}\cap dom(\textsf{MapletIndividual\_image\_Individual})$
 	\newline \hspace*{0.10in} \textbf{THEN} \textbf{\textsf{Property}:} $o\_Ind = o\_Ant \mapsto o\_Im$
   \newline \textbf{ELSE} \textbf{\textsf{LogicFormula}:}  $o\_Ant \mapsto o\_Im  \in o\_AS$\footnote{Following the variability status of \textit{o\_AS}, this predicate can be a property or an invariant}
   	\\
\hline
16 & \textbf{Variable individual initialisation} &
Ind Init CO Init\_ant Init\_im & $Ind \in \textsf{Individual}\cap dom(Individual\_name)$ 
\newline\newline  $Individual\_isVariable(Ind) = TRUE$ 
\newline\newline $o\_Ind \in \textsf{Variable}$  
 \newline\newline $CO = \textsf{Individual\_individualOf\_Concept}(Ind)$  
   \newline\newline $o\_CO \in \textsf{Set} \cup  \textsf{Constant} \cup \textsf{Variable}$
\newline\newline  $Ind \notin dom(\textsf{Individual\_initialValue\_individual}) \newline \vee (\textsf{Individual\_initialValue\_individual}(Ind) = Init \newline \wedge ((Init \notin dom(\textsf{Individual\_name}) \newline \wedge Init\_ant = \textsf{MapletIndividual\_antecedent\_Individual}(Init) \newline \wedge Init\_im = \textsf{MapletIndividual\_image\_Individual}(Init) \newline \wedge \{Init\_ant, Init\_im\} \subseteq \textsf{Constant} \cup \textsf{Variable}) \newline \vee o\_Init \in \textsf{Constant} \cup \textsf{Variable}))$   
&  & 
\textbf{IF}  $Ind \notin dom(\textsf{Individual\_initialValue\_individual})$
 \newline \textbf{THEN} $o\_Ind ::  o\_CO$ 
  \newline \textbf{ELSE} 
  \newline \hspace*{0.20in} \textbf{IF}  $Init \notin dom(\textsf{Individual\_name})$
 \newline \hspace*{0.20in} \textbf{THEN} \textbf{\textsf{Initialisation}:}  $o\_Ind :=   o\_Ant \mapsto o\_Im$
  \newline \hspace*{0.20in}  \textbf{ELSE} \textbf{\textsf{Initialisation}:} $o\_Ind :=  o\_Init$
\\   	
\hline
17 & \textbf{Variable concept initialisation} & CO $(I_j)_{j \in 1..n}$
&
$CO \in dom(\textsf{Concept})$
\newline\newline  $\textsf{Concept\_isVariable}(CO) =  TRUE$
\newline\newline $\forall j \in 1..n, I_j \in \textsf{Individual} \newline \wedge \textsf{Individual\_individualOf\_Concept}(I_j) = CO \newline \wedge \textsf{Individual\_isVariable}(I_j)=FALSE$
\newline\newline $o\_CO \in  \textsf{Variable}$
\newline\newline $\forall j \in 1..n, o\_I_j \in  o\_CO$
&  &
 \textbf{\textsf{Initialisation}:} $o\_CO := (o\_I_j)_{j \in 1..n}$\footnote{If $\exists j \in 1..n. I_j \notin dom(\textsf{Individual\_name})$ then $o\_I_j$ must be replaced by $o\_I_{j}\_Ant \mapsto o\_I_{j}\_Im$ as in the previous rule}
\\
\hline
18 & \textbf{Association transitivity} &
AS & $AS \in \textsf{Association}$ 
 \newline $\textsf{Association\_isTransitive}(AS) = TRUE$
 \newline $o\_AS \in \textsf{Constant} \cup \textsf{Variable}$
&  & \textbf{\textsf{LogicFormula}:} $(o\_AS ~ ; ~ o\_AS) \subseteq o\_AS$\\
\hline
19 & \textbf{Association symmetry} &
AS & $AS \in \textsf{Association}$ 
 \newline $\textsf{Association\_isSymmetric}(AS) = TRUE$
 \newline $o\_AS \in \textsf{Constant} \cup \textsf{Variable}$
&  & \textbf{\textsf{LogicFormula}:} $o\_AS^{-1}  = o\_AS$\\
\hline
20 & \textbf{Association asymmetry} &
AS CO & 
$AS \in \textsf{Association}$ 
 \newline $\textsf{Association\_isSymmetric}(AS) = TRUE$
 \newline $o\_AS \in \textsf{Constant} \cup \textsf{Variable}$
 \newline $\textsf{Association\_domain\_Concept}(AS) = CO$
 \newline $o\_CO \in \textsf{Set} \cup \textsf{Constant} \cup \textsf{Variable}$ 
&  & \textbf{\textsf{LogicFormula}:} $(o\_AS^{-1}  \cap o\_AS) \subseteq id(o\_CO)$\\
\hline
21 & \textbf{Association reflexivity} &
AS CO & 
$AS \in \textsf{Association}$ 
 \newline $\textsf{Association\_isReflexive}(AS) = TRUE$
 \newline $o\_AS \in \textsf{Constant} \cup \textsf{Variable}$
 \newline $\textsf{Association\_domain\_Concept}(AS) = CO$
 \newline $o\_CO \in \textsf{Set} \cup \textsf{Constant} \cup \textsf{Variable}$ 
&  & \textbf{\textsf{LogicFormula}:} $id(o\_CO) \subseteq o\_AS$\\
\hline
22 & \textbf{Association irreflexivity} &
AS CO & 
$AS \in \textsf{Association}$ 
 \newline $\textsf{Association\_isIrreflexive}(AS) = TRUE$
 \newline $o\_AS \in \textsf{Constant} \cup \textsf{Variable}$
 \newline $\textsf{Association\_domain\_Concept}(AS) = CO$
 \newline $o\_CO \in \textsf{Set} \cup \textsf{Constant} \cup \textsf{Variable}$ 
&  & \textbf{\textsf{LogicFormula}:} $id(o\_CO) \cap o\_AS = \emptyset$\\
\hline
\end{longtable}
\end{center}

\end{scriptsize}

Each logical formula  is translated with the definition of a \textit{B System} logic formula corresponding to its assertion.
Since both languages use first-order logic notations, the translation  is limited  to a syntactic rewriting.

\section{Updates in Back Propagation Rules from B System Specifications to Domain Models}

We choose to support only the most repetitive additions that can be performed within the formal specification,   the domain model remaining the one to be updated in case of any major changes such as the addition or the deletion of a refinement level. 
 Table \ref{tableau_recapitulatif_back_propagation} summarises the  most relevant back propagation rules.  
 Each rule defines its inputs (elements added to the \textit{B System} specification) and  constraints that each input must fulfill.
 It also defines its outputs (elements introduced within  domain models as a result of the application of the rule) and  their respective constraints.
It should be noted that for an element \textit{b\_x} of the \textit{B System} specification, \textit{o\_x} designates the domain model element corresponding to \textit{b\_x}. In addition, when used, qualifier \textit{abstract}  denotes "without parent".
	
\begin{scriptsize}
\begin{center}
\begin{longtable}{|p{.1cm}|p{2.2cm}|p{0.9cm}|p{6.cm}|p{0.9cm}|p{5.7cm}|}
\caption{\label{tableau_recapitulatif_back_propagation} back propagation rules in case of addition of an element in the \textit{B System} specification}\\
\hline
& &\multicolumn{2}{|c|}{\textbf{B System}} & \multicolumn{2}{|c|}{\textbf{Domain Model}} \\
\hline
& \textbf{Addition Of} &\textbf{Input} & \textbf{Constraint} & \textbf{Output} & \textbf{Constraint} \\
\hline
1 & \textbf{Abstract set}  & b\_CO & $b\_CO \in{} \textsf{AbstractSet} $ & o\_CO & $o\_CO \in \textsf{Concept}$
\\
\hline
2 & \textbf{Abstract enumeration}  & b\_CO $(b\_I_j)_{j \in 1..n}$  & $b\_CO \in{} \textsf{EnumeratedSet} $
\newline\newline $\forall j \in 1..n, b\_I_j \in \textsf{SetItem} \newline \wedge \textsf{SetItem\_itemOf\_EnumeratedSet}(b\_I_j) = b\_CO$ & o\_CO $(o\_I_j)_{j \in 1..n}$ & $o\_CO \in \textsf{Concept}$
\newline\newline $\textsf{Concept\_isEnumeration}(o\_CO) =  TRUE$
\newline\newline $\forall j \in 1..n, o\_I_j \in \textsf{Individual} \newline \wedge \textsf{Individual\_individualOf\_Concept}(o\_I_j) = o\_CO$
\\
\hline
 3 & \textbf{Set item} & b\_elt b\_ES & $b\_elt \in{} \textsf{SetItem}$
\newline $b\_ES = \textsf{SetItem\_itemOf\_EnumeratedSet}(b\_elt)$
\newline $o\_ES \in  \textsf{Concept}$  & o\_elt & 
$o\_elt \in \textsf{Individual} $ \newline
$\textsf{Individual\_individualOf\_Concept}(o\_elt) = o\_ES$
 \\
 \hline
4 & \textbf{Constant typed as subset of the correspondent of a concept} & b\_CO b\_PCO &$b\_CO \in \textsf{Constant} $\newline $b\_PCO \in \textsf{AbstractSet} \cup \textsf{Constant}$ \newline$b\_CO \subseteq b\_PCO$ \newline $o\_PCO \in \textsf{Concept}$  & o\_CO & 
$o\_CO \in \textsf{Concept}$ \newline
$\textsf{Concept\_parent\_Concept}(o\_CO) = o\_PCO$ 
 \\ 
  \hline
5 & \textbf{Constant typed as item of the correspondent of a concept} & b\_elt b\_CO 
&$b\_elt \in \textsf{Constant}$ 
\newline $b\_CO \in \textsf{AbstractSet} \cup \textsf{Constant}$ 
\newline $b\_elt \in  b\_CO$
 \newline $o\_CO \in \textsf{Concept}$  & o\_elt & 
$o\_elt \in \textsf{Individual} $ \newline
$\textsf{Individual\_individualOf\_Concept}(o\_elt) = o\_CO$ 
 \\ 
  \hline
6 & \textbf{Variable typed as subset of the correspondent of a concept} & b\_CO b\_PCO &$b\_CO \in \textsf{Variable} $\newline $b\_PCO \in \textsf{AbstractSet} \cup \textsf{Constant} \cup \textsf{Variable}$ \newline$b\_CO \subseteq b\_PCO$ \newline $o\_PCO \in \textsf{Concept}$  & o\_CO & 
$o\_CO \in \textsf{Concept}$ \newline
$\textsf{Concept\_parent\_Concept}(o\_CO) = o\_PCO$ 
\newline $\textsf{Concept\_isVariable}(CO) =  TRUE$
 \\ 
  \hline
7 & \textbf{Variable typed as item of the correspondent of a concept} & b\_elt b\_CO  
&$b\_elt \in \textsf{Variable}$ 
\newline $b\_CO \in \textsf{AbstractSet} \cup \textsf{Constant} \cup \textsf{Variable}$ 
\newline $b\_elt \in  b\_CO$
 \newline $o\_CO \in \textsf{Concept}$  & o\_elt & 
$o\_elt \in \textsf{Individual} $ \newline
$\textsf{Individual\_individualOf\_Concept}(o\_elt) = o\_CO$ 
\newline $\textsf{Individual\_isVariable}(o\_elt) =  TRUE$
 \\ 
  \hline
8 &  \textbf{Constant  typed as a relation} & b\_AS b\_CO1 b\_CO2 &$b\_AS \in \textsf{Constant}$ \newline $\{b\_CO1, b\_CO2\} \subset \textsf{AbstractSet} \cup \textsf{Constant} $ \newline$b\_AS \in  b\_CO1 \leftrightarrow b\_CO2$
\newline $\{o\_CO1, o\_CO2\} \subset \textsf{Concept}$   & o\_AS & 
$o\_AS \in \textsf{Association} $ \newline
$\textsf{Association\_domain\_Concept}(o\_AS) =  o\_CO1$
\newline
$\textsf{Association\_range\_Concept}(o\_AS) =  o\_CO2$
\newline
{ As usual, the cardinalities of \textit{o\_AS} are set according to the type of \textit{b\_AS} (\textit{function},  \textit{injection}, ...).}
 \\ 
  \hline
9 &  \textbf{Variable typed as a relation} & b\_AS b\_CO1 b\_CO2 &$b\_AS \in \textsf{Variable}$ \newline $\{b\_CO1, b\_CO2\} \subset \textsf{AbstractSet} \cup \textsf{Constant} \cup \textsf{Variable}$  \newline$b\_AS \in  b\_CO1 \leftrightarrow b\_CO2$
\newline $\{o\_CO1, o\_CO2\} \subset \textsf{Concept}$   & o\_AS & 
$o\_AS \in \textsf{Association} $ \newline
$\textsf{Association\_domain\_Concept}(o\_AS) =  o\_CO1$
\newline
$\textsf{Association\_range\_Concept}(o\_AS) =  o\_CO2$
\newline
$\textsf{Association\_isVariable}(o\_AS) =  TRUE$
\newline
{ As usual, the cardinalities of \textit{o\_AS} are set according to the type of \textit{b\_AS} (\textit{function},  \textit{injection}, ...).}
 \\ 
  \hline
10 &  \textbf{Constant  typed as a maplet} & b\_elt b\_ant b\_im &$b\_elt \in \textsf{Constant}$ \newline $\{b\_ant, b\_im\} \subset  \textsf{Constant} $ \newline $b\_elt =  b\_ant \mapsto b\_im$
\newline $\{o\_ant, o\_im\} \subset \textsf{Individual}$   & o\_elt & 
$o\_elt \in \textsf{Individual} $ \newline
$\textsf{MapletIndividual\_antecedent\_Individual}(o\_elt) =  o\_ant$
\newline
$\textsf{MapletIndividual\_image\_Individual}(o\_elt) =  b\_im$
 \\ 
 \hline
11 &  \textbf{Variable  typed as a maplet} & b\_elt b\_ant b\_im &$b\_elt \in \textsf{Variable}$ \newline $\{b\_ant, b\_im\} \subset  \textsf{Constant} \cup \textsf{Variable}$ \newline $b\_elt =  b\_ant \mapsto b\_im$
\newline $\{o\_ant, o\_im\} \subset \textsf{Individual}$   & o\_elt & 
$o\_elt \in \textsf{Individual} $ \newline
$\textsf{MapletIndividual\_antecedent\_Individual}(o\_elt) =  o\_ant$
\newline
$\textsf{MapletIndividual\_image\_Individual}(o\_elt) =  b\_im$
\newline $\textsf{Individual\_isVariable}(o\_elt) =  TRUE$
 \\ 
 \hline
12 & \textbf{Variable initialised to the correspondent of an individual} & b\_elt b\_init  
&$b\_elt \in \textsf{Variable}$ 
\newline $b\_init \in  \textsf{Constant}$ 
\newline  \textbf{\textsf{Initialisation}:} $b\_elt :=  b\_init$
 \newline $\{o\_init, o\_elt\}  \subseteq \textsf{Individual}$  &  & 
$\textsf{Individual\_initialValue\_Individual}(o\_elt) = o\_init$ 
 \\ 
\hline

\end{longtable}
\end{center}
\end{scriptsize}

The addition of a non typing logic formula (logic formula that does not contribute to the definition of the type of a formal element) in the \textit{B System} specification is propagated through the definition of the same formula  in the corresponding domain model, since both languages use first-order logic notations. This back propagation is limited  to a syntactic translation.

A fresh \textit{B System} constant or variable \textit{b\_x} is defined within the domain model, by default, as a defined concept (instance of \textsf{DefinedConcept}), until a typing \textit{B System} logical formula is introduced (subset of the correspondence of a concept, relation, item of the correspondence of a concept or  maplet).
The  concept \textit{b\_x} is defined with correspondence of \textit{B System}  logical formulas where \textit{b\_x} appears: there must be at least one.

\section*{Acknowledgment}
This work is carried out within the framework of the  \textit{FORMOSE} project \citep{anr_formose_reference_link} funded by the French National Research Agency (ANR).
It  is also partly supported by the Natural Sciences and Engineering Research Council of Canada  (NSERC).




\bibliographystyle{elsarticle-num}
\bibliography{references}

\end{document}